\documentclass[mathleft]{an}
\usepackage{graphicx}
\usepackage{times}
\usepackage{hyperref}
\hypersetup{colorlinks,
 citecolor=blue,
 linkcolor=blue}

\begin{document}

\Pagespan{789}{}
\Yearpublication{2009}%
\Yearsubmission{2009}%
\Month{11}%
\Volume{999}%
\Issue{88}%

\title{Stellar populations of seven early-type dwarf galaxies  and their nuclei\,\thanks{Based on observations made at the European Southern Observatory, Chile(ESO Programme ID 078.B-0178(A))}}

\author{S. Paudel\inst{1}\fnmsep\thanks{Corresponding author:
  \email{sjy@x-astro.net}\newline}
\and  T. Lisker\inst{1}
}
\titlerunning{Stellar populations of dEs}
\authorrunning{S. Paudel \& T. Lisker}
\institute{
Astronomisches Rechen-Institut, Zentrum f\"ur Astronomie der Universit\"at
  Heidelberg, 69120 Heidelberg, Germany}

\received{August 2009}
\publonline{later}

\keywords{galaxies:dwarf -- galaxies: evolution -- galaxies: formation -- galaxies: statistics}

\abstract{%
 Dwarf galaxies are  the numerically dominating population in the dense regions of the universe. Although they seem to be simple systems at first view, the stellar populations of dwarf elliptical galaxies (dEs) might be fairly complex. Nucleated dEs are of particular interest, since a number of objects exhibit different stellar populations in their nuclei and host galaxy. We present stellar population parameters obtained from integrated optical spectra using a Lick index analysis of seven nucleated dwarf elliptical galaxies and their nuclei. After subtracting the scaled galaxy spectra from the nucleus spectra, we compared them with one another and explore their stellar populations.  As a preliminary result, we find that the luminosity weighted ages of the nuclei slightly lower than those of galaxies, however, we do not see any significant difference in metallicity of the host galaxies and their nuclei.}

\maketitle

\section{Introduction}
Though diffuse elliptical galaxies represent the majority of the galaxy populations in dense regions of the nearby Universe like rich clusters, their origin and evolution remain still a matter of debate. Several recent and past studies show that these galaxies exhibit a great variety of kinematics and stellar population properties (Michielsen et al. 2008; van Zee, Skillman \& Haynes 2004; Poggianti et al. 2001; Mateo1998;) posing a number of questions to the current understanding of the external and internal evolution of dwarf elliptical galaxies. A recent systematic study by Lisker et al. 2008, 07, 06 of dEs in the Virgo cluster revealed a striking heterogeneity of this galaxy class. They found different subclasses, with significantly different shapes, colors, and spatial distributions. Those dEs with a disk component have a flat shape and are predominantly found in the outskirts of the cluster, suggesting that they -- or their progenitors -- might have just recently fallen into the cluster environment. In contrast, dEs with a compact stellar nucleus follow the classical picture of dwarf ellipticals: they are spheroidal objects that are preferentially found in the dense cluster center. Interestingly, though, the majority of dEs with disks also host a nucleus in their center. This would not seem too surprising if one assumed that the nuclei were already in place before the galaxy received its present-day configuration. However, several proposed mechanisms for nucleus formation are based on \emph{late} nucleus formation, e.g. Van den Bergh (1986) proposed that the nuclei of dE,N could have formed from gas that sank to the centre of the more slowly rotating objects. Furthermore, rounder dEs with compact nuclei are predominantly present in highly dense environments like the centre of galaxy clusters. The pressure from the surrounding inter-galactic medium may allow dwarf galaxies to retain their gas during star formation and produce multiple generations of stars (Babul 1992, Silk et al. 1987), forming nuclei in the process. In both proposed scenarios, the stars constituting the nuclei are formed late, out of gas within the galaxy. In contrast, Oh \& Lin 2000 suggested that the nuclei might have formed from globular clusters that migrated into the galaxy center, thus meaning that the nucleus' stars were formed separately and probably earlier than most stars of the host galaxy. Large globular clusters or super star clusters are candidates for becoming such nuclei of dEs. Given these different possibilities for the formation process of nuclei, and also of dEs itself, can the nuclei thus tell us something about the formation history of their host galaxies?\\
\\
Motivated by these reasons for the importance of investigating stellar population differences between the nuclei and the host dEs outside of the central region, we have studied seven nucleated dEs of the Virgo cluster, using medium resolution spectroscopy. \\

\section{Sample and Observation}

The sample consists of seven early-type nucleated dwarf galaxies of the Virgo cluster, having different substructure or color characteristics: one has a blue central region ( dE(bc) ), two show faint signatures of disks (dE(di) ), and four others have no such features and are simply
nucleated ( dE(N) ), according to Lisker et al. 2007. The observations were carried out over six half nights between March 16 to 18, 2007, with the FORS2 instrument mounted on UT1 at ESO VLT, using the multi-object spectroscopy (MXU) mode. Integration times varied between 10 and 30 minutes, depending on the surface brightness of each galaxy. The galaxy properties and the final signal-to-noise ratio of the extracted spectra are listed in Table \ref{stb1}. The spectra are flux calibrated using spectrophotometric standard stars, which were observed during the run. A slit width of 1" and length of 40" were used, with the 300V grism providing a dispersion of 3.36 \AA{} / pixel. This setup provided a spectral resolution, as measured from the FWHM of the arc lines, of $\sim$11 \AA{} at $\sim$5000 \AA, which is somewhat below the Lick resolution ($\sim$8.4 \AA{} at 5000 \AA). \\
\begin{table}
\begin{center}
\begin{tabular}{|c|c|r|r|r|r|}
\hline
Name	&	RV	&	Mag	&	Type 	&	S/N gal.	&	S/N nuc	\\
\hline
VCC0308		&	1850$\pm$39 	&	13.19	&	bc  &	     34	&	33	\\
VCC0216        &	1281$\pm$26 	&	14.35	&	di    &    34	&	44	\\
VCC0490		&	1267$\pm$12 	&	13.80	&	di  &	     27	&	32	\\
VCC0929		 &	0910$\pm$10 	&	12.60	&	n    &	43	&	53	\\
VCC1254		 &	1278$\pm$18 	&	14.27	&	n  &	     32	&	65	\\
VCC1348		&	1968$\pm$25 	&	14.33	&	n   &	     27	&	42	\\
VCC1945		&	1619$\pm$10	&	14.50	&	n	    &	37	&	30	 \\
\hline
\end{tabular}
\end{center}
\caption{The galaxies are identified by their number in the Virgo cluster catalog (VCC, Binggeli et al. 1985)}
\label{stb1}
\end{table}\\
We integrated over the interval 2"$<$r$<$7" along the spatial direction, to extract the spectrum of the underlying host galaxy, thereby considering that the nucleus light is negligible beyond the maximum size of the seeing disk (1.5") that occured for our sample. Since we know that the nucleus light sits on top of the underlying galaxy, we subtracted the galaxy light from the central nucleus spectrum. This process is illustrated in Fig. 1: first, we fitted the galaxy light to a smooth exponential profile beyond 2" from the centre, then we extrapolated the galaxy light inwards to the centre, and subtract this central galaxy light from the total light. Finally, we integrate over the central 3 pixels (0.75") along the spatial direction to extract one-dimensional nucleus spectra.
\begin{figure} 
 \includegraphics[width=8cm,height=9cm]{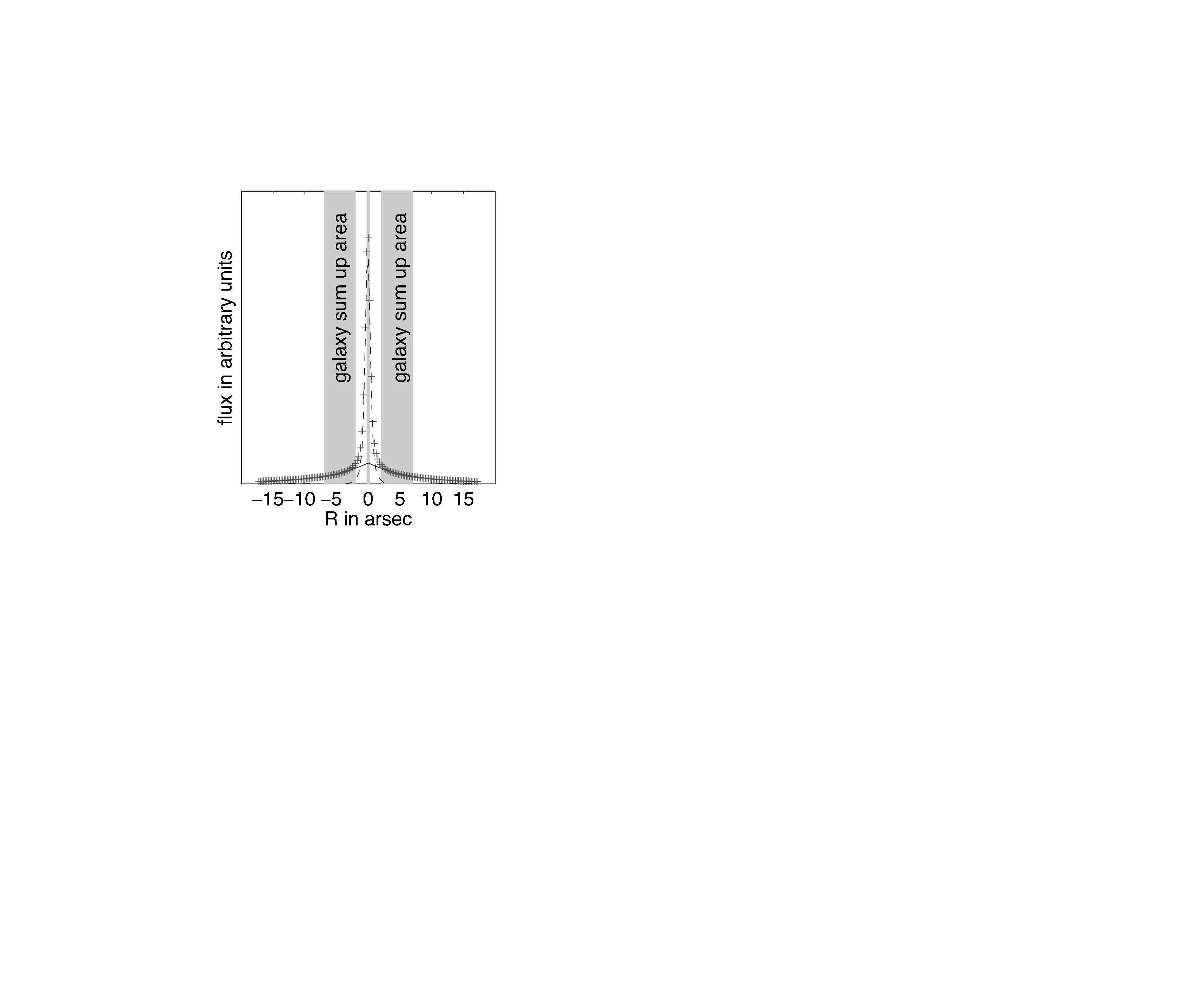}
 \caption{Schematic view of the fitting of the light profile of the galaxy and nucleus. The crosses represent the distribution of total light (i.e galaxy + nucleus) and solid line represents the exponentially fitted light profile of the galaxy. The dashed line is the residual nucleus flux after the subtraction of the galaxy light.}
 \label{sef}
\end{figure}

\section{Result}
\begin{figure}
 \includegraphics[width=8cm,height=7cm]{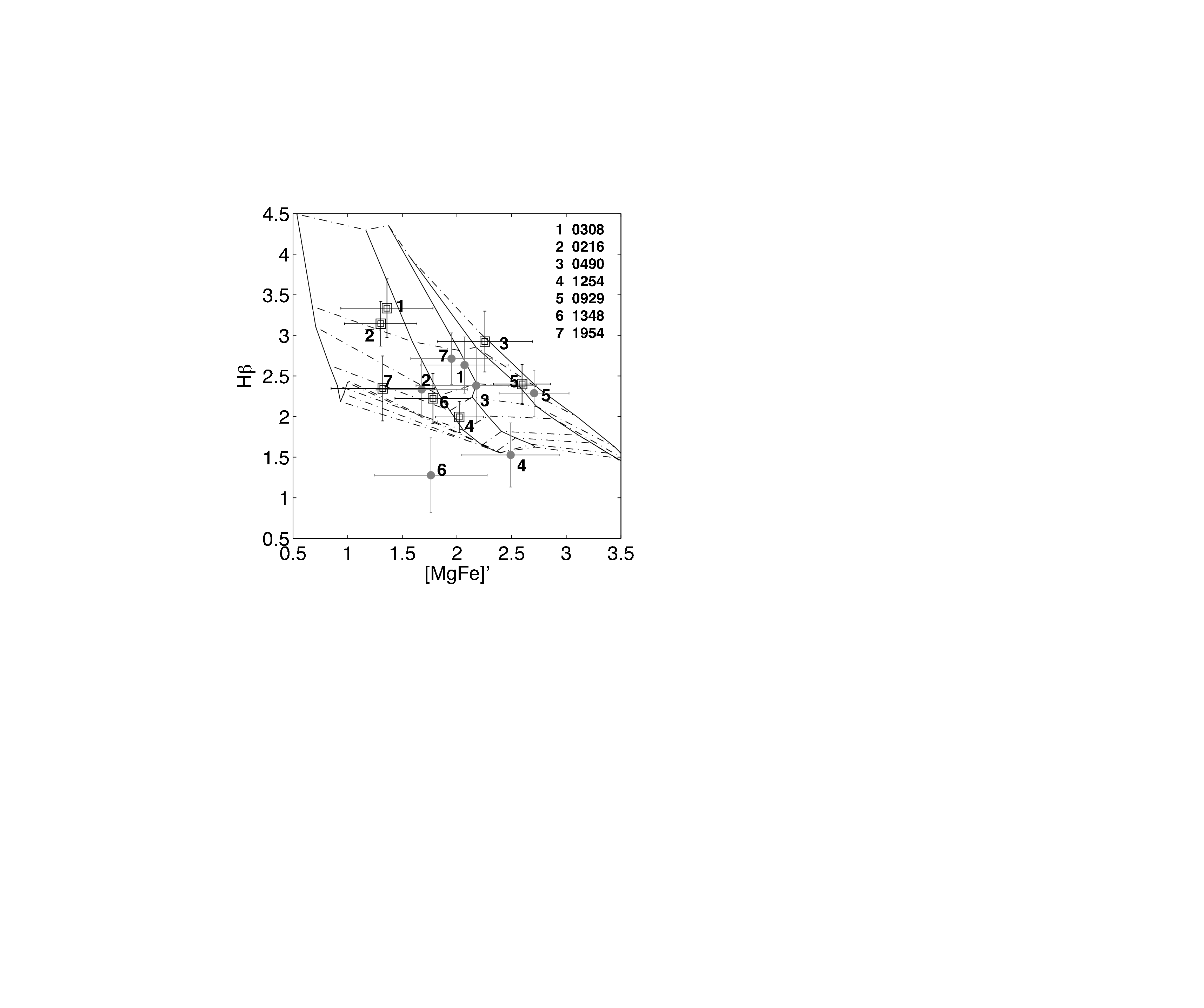}
\caption{The age-sensitive index H$\beta$ as a function of the metallicity-sensitive index [MgFe]. Overplotted are the stellar population models of BC03. The solid lines are lines of constant age from 1 to 15 Gyr. The solid, almost vertical, lines stand for constant metallicity from [Z/H] = -1.64 to 0.55 dex for BC03. For both this plot and Fig. 3, the nuclei are shown with black squares, while the host galaxies are shown by filled gray circles.}
\label{hmfe}
\end{figure}

\begin{figure}
 \includegraphics[width=6cm,height=7cm]{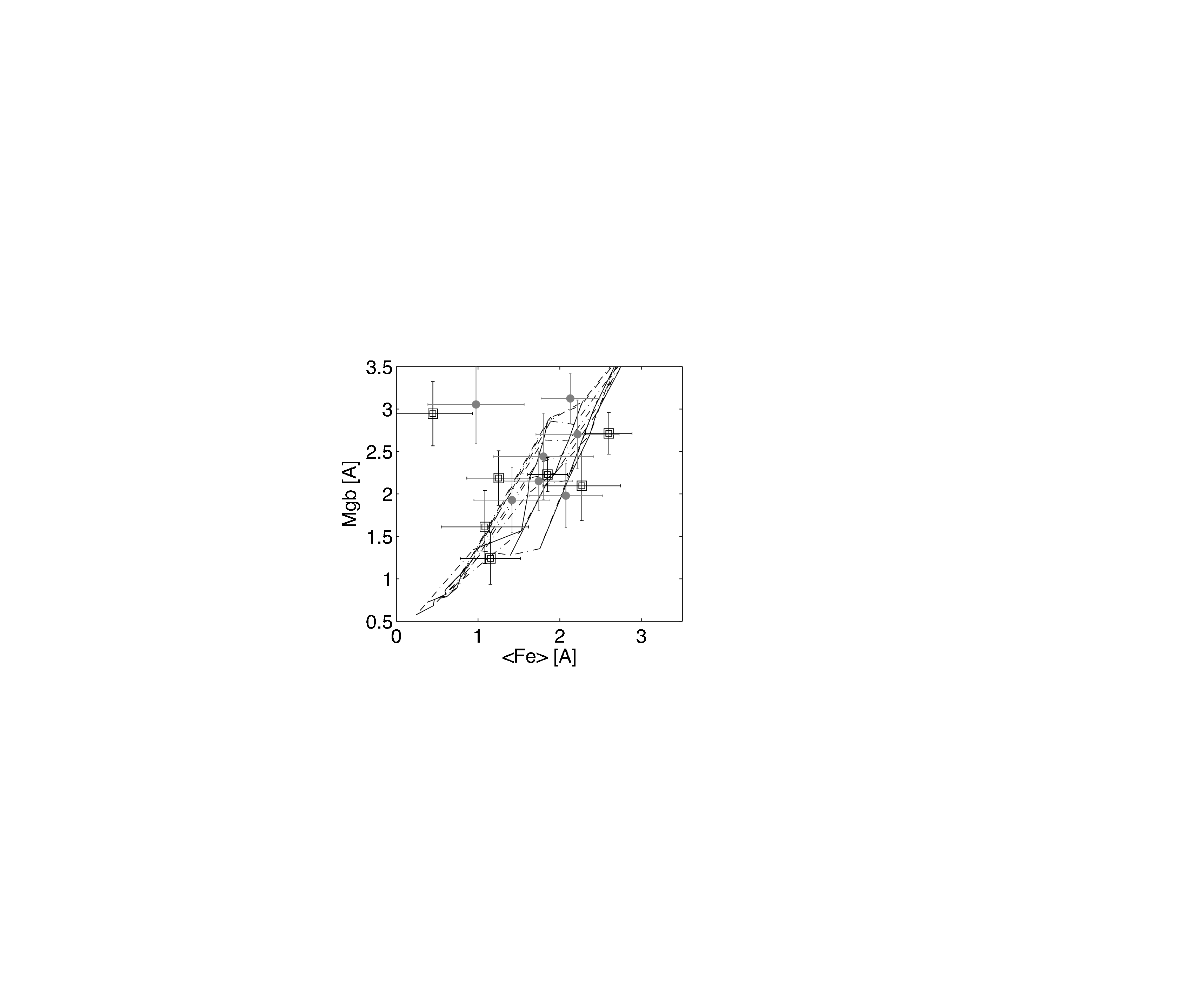}
 \label{mgfe}
 \caption{$<$Fe$>$ Vs Mgb diagram. }
 \end{figure}

The luminosity-weigthed age and metallicity, can be inferred from a comparison of selected line-strength indices with models of single stellar populations such as those of Bruzual \& Charlot 2003 (BC03 here after). In the Fig \ref{hmfe}, we show Lick/IDS index-index diagrams with SSP model grid taken from BC03 based on Padova 1994 isochrones. Note that, Our spectra are flux calibrated but the resolution is somewhat lower than model resolution, we therefore degrade the BC03 model spectra down to our resolution in order to match the resolution of the model and observed spectra. From H$\beta$ - [MgFe]' diagram in the Fig \ref{hmfe}, we can infer luminosity weighted ages and metallicity of nuclei and their host galaxies. Black squares represent the nuclei and filled gray circles the respective host galaxies. It is conspicuous that most of the nuclei are located upward of their respective host galaxies (see the number labels in the figure), indicating that their stellar population is younger than that of the host. Furthermore, the one dE with a blue central region, VCC 308, has the youngest nucleus, confirming the recent star-formation activity in its centre. In contrast to that, only one (VCC 1945) shows the opposite trend, namely that the galaxy is younger than its nucleus. We do not see any significant difference in metallicity between nuclei and host galaxies. Although the sample is quite small, we do not find any difference in the stellar populations of dEs and their nuclei with respect to different substructure properties.\\
\\
The $<$Fe$>$-Mgb diagram (Fig. \ref{mgfe}) gives an estimate of the $\alpha$/Fe element ratio, relative to solar values of the model. The almost all the dEs and and their nuclei are consistent with solar abundance, however one VCC1348 behave differently, lying in the region of super solar abundance.

\section{Conclusion}
Although the numbers are small, we find that dE nuclei are slightly younger than their host galaxies. 
Such a difference in the age of nuclei and galaxies supports the theory of late nucleus formation, in the sense that a non-nucleated dE formed first, and the nucleus formed subsequently within the dE due to secondary star formation (e.g. Bothun \& Mould 1988) or infalling gas (e.g. Silk et al. 1987) leading to continuous star formation activity in the nucleus.
\section{Acknowledgement}
We are supported within the framework of the Excellence Initiative by the German Research Foundation (DFG) through the Heidelberg Graduate School of Fundamental Physics (grant number GSC 129/1). S.P. acknowledges the support of the International Max Planck Research School (IMPRS) for Astronomy and Cosmic Physics at the University of Heidelberg.


\begin{thebibliography}{}
\bibitem{}Babul, A. \& Rees, M. 1992, MNRAS, 255, 346
 \bibitem{}Bothun G. D., Mould J. R., 1988, ApJ, 324, 123
 \bibitem{}Bruzual, G., \& Charlot, S. 2003, MNRAS, 344, 1000 (BC03)
\bibitem{}Lisker, T., Grebel, E., Binggeli, B., 2006a, AJ, 132, 497 
\bibitem{}Lisker, T., Glatt, K., Westera, P., Grebel, E., 2006b, AJ, 132, 2432 
\bibitem{}Lisker, T., Grebel, E. K., Binggeli, B., \& Glatt, K. 2007, ApJ, 660, 1186
\bibitem{}Lisker, T., Grebel, E., Binggeli, B., 2008, AJ, 135, 380
\bibitem{}Michielsen D., Boselli A., Conselice C. J., Toloba E., Whiley I. M., Arag on-Salamanca A., Balcells M., Cardiel N., Cenarro A. J., Gorgas J., Peletier R. F., Vazdekis A., 2008, MNRAS, 385, 1374 
\bibitem{}Mateo M. L., 1998, ARA\&A, 36, 435 
\bibitem{}Oh K. S, Lin D. N. C., 2000,ApJ, 543, 620
\bibitem{}Poggianti B. M., Bridges T. J., Mobasher B., Carter D.,  Doi M., Iye M., Kashikawa N., Komiyama Y., Okamura S., Sekiguchi M., Shimasaku K., Yagi M., Yasuda N., 2001,  ApJ, 562, 689
 \bibitem{}Silk J., Wyse R. F. G., \& Shields G. A., 1987, ApJ, 322, L59
\bibitem{}Van den Bergh S., 1986, AJ, 91, 271
\bibitem{}van Zee L., Skillman E. D., Haynes M. P., 2004, AJ, 128, 121  
\end{thebibliography}
\end{document}